\documentclass[11pt]{article}
\usepackage{setspace}
\usepackage{graphicx}  
\usepackage{array}
\usepackage{url} 
\usepackage[linesnumbered,ruled,vlined]{algorithm2e}

\SetKwInput{KwInput}{Input}                
\SetKwInput{KwOutput}{Output}              

\begin{document}
\title{Load Balanced Semantic Aware Distributed RDF Graph }

\author{
Ami Pandat\\
       Distributed Databases Group\\
       DAIICT, Gandhinagar\\
       \texttt{202021002@daiict.ac.in}
       \and
       Nidhi Gupta\\
       Distributed Databases Group\\
       DAIICT, Gandhinagar\\
       \texttt{201601048@daiict.ac.in}
       \and
       Minal Bhise\\
       Distributed Databases Group\\
       DAIICT, Gandhinagar\\
       \texttt{minal\_bhise@daiict.ac.in}}

\date{}   

\maketitle
~\\
\begin{abstract}
The modern day semantic applications store data as Resource Description Framework (RDF) data.Due to Proliferation of RDF Data, the efficient management of huge RDF data has become essential. A number of approaches pertaining to both relational and graph-based have been devised to handle this huge data. As the relational approach suffers from query joins, we propose a semantic aware graph based partitioning method. The partitioned fragments are further allocated in a load balanced way. For efficient query processing, partial replication is implemented. It reduces Inter node Communication thereby accelerating queries on distributed RDF Graph. This approach has been demonstrated in two phases partitioning and Distribution of Linked Observation Data (LOD). The time complexity for partitioning and distribution of Load Balanced Semantic Aware RDF Graph (LBSD) is O(n) where n is the number of triples which is demonstrated by linear increment in algorithm execution time (AET) for LOD data scaled from 1x to 5x.  LBSD has been found to behave well till 4x. LBSD is compared with the state of the  art relational and graph-based partitioning techniques. LBSD records 71\% QET gain when averaged over all the four query types. For most frequent query types, Linear and Star, on an average 65\% QET gain is recorded over original configuration for scaling experiments. The optimal replication level has been found to be 12\% of  original data.
\end{abstract}
~\\


{\bf Topics}    DB Systems and Applications,    Distributed Databases,  Sensor Data Management
~\\

\section{Introduction}

A W3C standard, Resource Description Framework (RDF) is a foundation of semantic web and used to model web objects. An RDF dataset comprises triples in the form of (subject, property, and object). It can be readily comprehended as a graph, where subjects and objects are vertices joined by labeled relationships i.e., edge.  It is however now being used in a broader context. Bio2RDF\cite{bio2} data collection is used by biologists to store their experimental results in RDF triples to support structural queries and communicate among themselves. Similarly, DBpedia\cite{dbp} extracts information from Wikipedia and stores it as RDF data. W3C offers a structured query language, SPARQL to retrieve and manage the RDF datasets. Finding an answer to the SPARQL query requires finding a match of the subquery graph in the entire RDF graph. As the RDF data is gaining acceptance widely, RDF dataset sizes are moving from a centralized system to distributed system.

There are two techniques for RDF data management: relational and graph-based. In the relational method, data is kept in the form of multiple tables. To find an answer to a query, one needs to extract that information from various tables by applying the join operation. On the other hand, in the graph-based technique data is represented in the form of vertices and edges. Semantic partitioning \cite{semantic} is one of the graph partitioning technique, implemented for a centralized system using page-rank algorithms. To work towards building efficient partitioning and distribution algorithms, there are many state of the art available. Some of the partitioning algorithms use the query workload to identify the parts of the RDF graph which are used frequently and keep these subgraphs at one site. While this approach works well for the systems in which the majority of queries follow the identified query patterns, it may not work as well in the systems where new queries do not correlate with the existing workload. The configuring system that doesn’t use workload information is desirable. Instead, if we use the semantics of RDF to partition the data, algorithm execution time would be much lower and query execution time for new queries would either be the same or better than the workload aware methods. Semanticity of RDF data refers to the format of triples in a Turtle or N-Quad RDF file. This triple data file can directly be used for partition and distribution using the fact that the edge is denoted by the equivalence of subject and object in two triples. Using this structure of triples, one can directly work on complexities that are based on the number of triples in a file. 

Reviewing such kind of aspects and agendas available in graph-based techniques, this research is designed to develop algorithms to partition data using semantic relation between vertices and distribute among several nodes. Load Balanced Semantic Aware Graph (LBSD) uses semantic partitioning, for the initial phase of partitioning. The system partitions data and makes clusters. At that point, it will disseminate applicable bunches (by semantic connection) among the given number of hubs. The fundamental reason to segment RDF information is to answer inquiries effectively in a lesser measure of time. To reduce inter node communication(INC) in distributed environment, partial replication \cite{icdcit} of data has been done. It is demonstrated by deciding how much amount of data should be replicated over every node to reduce INC. 

The rest of the paper is organized as follows: in the next section, we discuss related work regarding this research. In Section 3, we discuss the methodology used to implement this work. Section 4 describes the details of experiments and evaluation parameters. In Section 5, we discuss the results and comparison of the system with the state of the  art work, and then finally Section 6 states the conclusion.

\section{Related Work}

The present approaches for handling the huge RDF data can be classified into two categories; Relational and Graph-based approaches.

\subsection{Relational Approaches}
RDF triples can naturally be implemented as a single table with three columns specifically subject, predicate object. This table can have millions of triples. This approach aims to utilize the well developed techniques in conventional relational techniques for query processing, and storage of data. Research in relational techniques deals with the partitioning of RDF tables in such a way that there is a substantial decrease in the number of joins while answering a query. Property tables approach utilizes the repeated appearances of patterns and stores correlated properties in the same table. Class property table and clustered property table are two techniques in which the former defines various tables that contain a particular property value while the latter defines a table for a particular subject\cite{AbadiMMH07}.

DWAHP \cite{dwahp} is the relational technique partitions the data using workload aware approach using n-hops property reachability matrix. Clustering of Relational data in distributed databases for medical information is discussed in \cite{ideas20} which is also similar kind of the state of the  art work for relational systems. It uses Horizontal Fragmentation for the implementation. This technique is implemented for relational approach and this research LBSD discusses the same for graph-based approach. The relational approach for SPARQL-based query known as Direct relational mappings in which a SPARQL query can be translated to SQL query for given data in the form of the triple \cite{grphbsdrdf}. Another technique is single table extensive indexing which is used to develop native storage systems that allow extensive indexing of the triple table. e.g. Hexastore and 3X \cite{survey}. SIVP \cite{bhavikshah_index} proposes Structure Indexed Vertical Partitioning which combines structure indexing and vertical partitioning to store and query RDF data for interactive semantic web applications. It presents five metrics to measure and analyze the performance of the SIVP store. SIVP is better than vertical partitioning provided the extra time needed in SIVP, which consists of lookup time and merge time, is compensated by frequency. Above all are relational approaches which closely relate to LBSD in some or other way.

\subsection{Graph Based Approaches}

The graph-based technique eliminates query joins. It maintains the original representation of the RDF data and implements the semantics of RDF queries but it scales poorly. Several recent works deal with RDF graph partitioning. gStore\cite{grphbsdrdf} is a system designed to exploit the natural graphical structure of RDF triples. It also executes the queries using the subgraph matching approach. The Graph-based technique, Adaptive partitioning and Replication (APR) \cite{lenawiesecnf} works to partition query graphs using Workload information, and then it decides the benefit level to make a certain decision that how much data should be replicated in the given graph. 

Another approach is UniAdapt \cite{uniad}. This technique proposes a unified optimization approach that enables a distributed RDF Triple Store to adapt its RDF Storage layer by focusing on replication as well as main memory indexes. The final objective for this approach to decrease future query execution time. METIS \cite{metis} is one of the popular baselines for multiple works. \cite{warp} \cite{sn} \cite{partout}. APR \cite{lenawiesecnf} first partitions the graph using METIS and then uses a global query graph made using workload for replication.

The other approach uses the semantic properties of RDF data and proposes a Page Rank inspired algorithm to cluster the RDF data \cite{semantic}. This approach is implemented for centralized system whereas proposed technique LBSD inspired by the same but works for distributed systems. One more recent approach \cite{peng} uses the frequency of query patterns to partition the graph and proposes three methods of fragmentation. Other than relational and graph-based approach there are approaches which deal with index, dataset formats and storage structure. While partitioning and distributing data, the index of data fed to the system and the format of data are also key features. Several partitioning techniques available to handle query workload for static partitioning, which turns into the result that 40\% query remains unanswered \cite{anubhadi}. These types of shortcomings are resolved in \cite{wise}, which handles dynamic ranged partitioning using workload information. 

To address limitations observed in above mentioned work, LBSD is developed to support semantic aware partitioning in a distributed environment which has two phases: 1. Semantic aware partitioning 2. Distribution using partial replication. It aims to reduce the communication cost during SPARQL query processing. It adaptively maintains some frequent access patterns (FAPs) to reflect the characteristics of the workload while ensuring the data integrity and approximation ratio. To reduce INC, data should be replicated among all local nodes by its semantic relation and for that, a partial replication technique can be used. The partial replication technique decides the replication level using certain criteria and replicates the vertices which are most frequently used or most relevant. The  Partial Replication technique \cite{icdcit}, finds the most frequent pattern and store it into a heat map table. Using this information it decides the replication level. LBSD uses the similar technique for graph based approach using Centrality concept.

\section{Research Methodology}
\begin{figure}[h!]
    \centering
    \includegraphics[width=0.6\columnwidth]{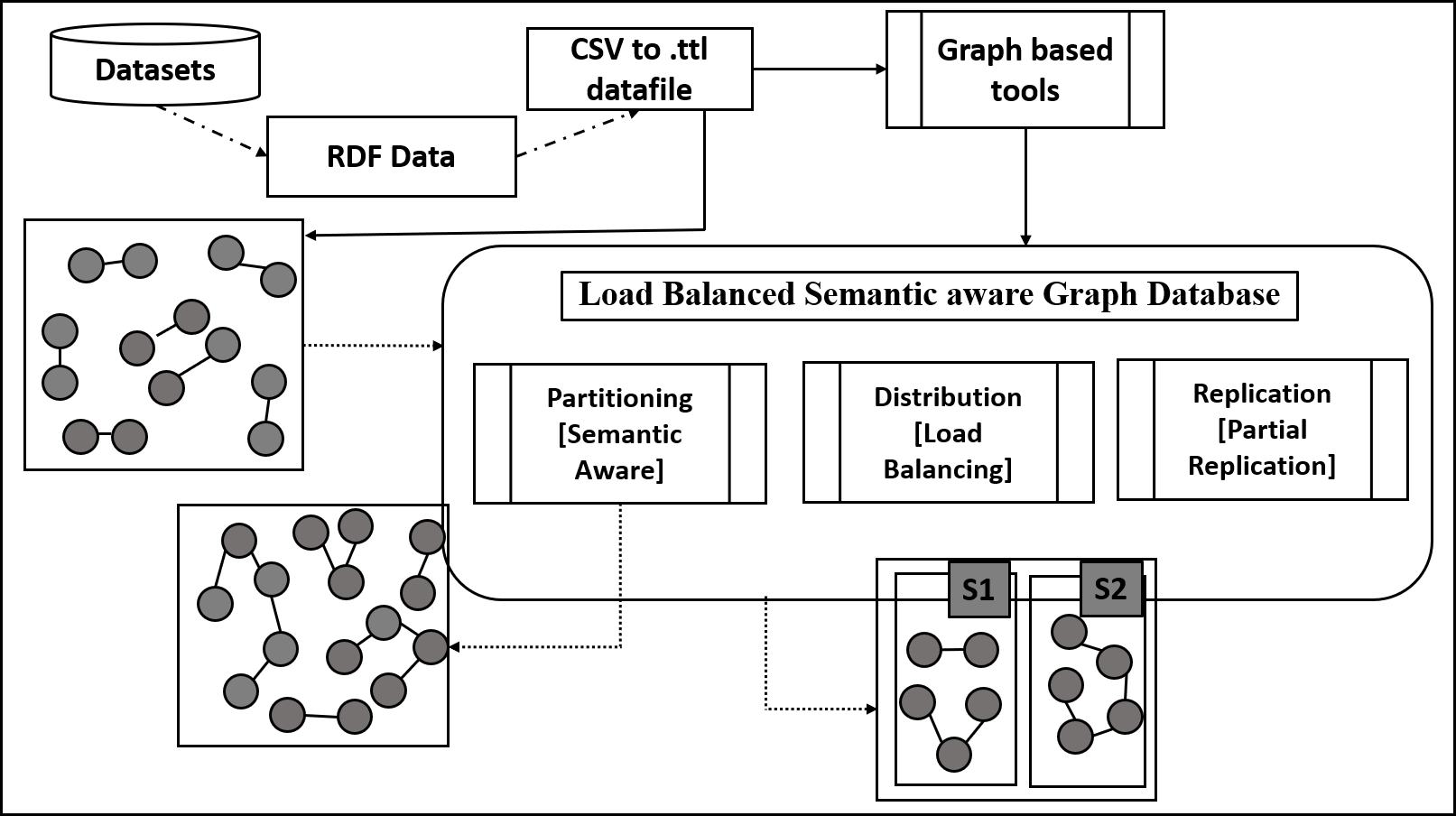}
    \caption{Block diagram for LBSD}
    \label{fig:datastruc}
\end{figure}

The LBSD aims to distribute RDF data using graph passed approach over available nodes to reduce inter-node communication (INC). The methodology divided into two phases. First Phase is Semantic aware Partitioning of RDF Data which consists of two algorithms. Algorithm 1 is used for extraction of popular nodes and algorithm 2 is used for partitioning.  The Second Phase is Distribution of RDF Data, includes algorithm 3 and algorithm 4 for distribution and replication respectively. Figure \ref{fig:datastruc} depicts the same. As shown in figure \ref{fig:datastruc}, first available datasets of RDF Data will be transformed from CSV to ttl datafile to set input into graph-based tools. The .ttl datafile will be as tripled data which then will be fragmented and distributed in upcoming phases.

\subsection{Partitioning of RDF Graph}
Our aim for designing a fragmentation algorithm is to reduce INC, especially for linear and star queries. For example, social media data may have frequent star queries to get the friends of a person. RDF data has an advantage because it represents the data in the form of triple $<subject, predicate, object>$. First, we need to find out the subjects which have many outgoing degrees. If we put these popular subjects at different nodes, then we can get rid of INC for star queries. 

Each of these popular subjects is known as the master nodes. We can do this by sending the incoming query to the node in which the corresponding subject resides. Suppose there are k fragments, then we need to find k subjects with the highest number of outgoing links. Algorithm 1 lists out the steps needed to do that.  Since algorithm 1 goes through all the triples only once, it has a linear time complexity O(n) where n is the number of triples. Sorting HashMap would take O(m*logm) where m is the distinct number of subjects. The number of triples is always much greater than the number of distinct subjects, we can safely ignore it.

\begin{algorithm}[H]
\SetKwFunction{FEXT}{EXTRACTION}
\DontPrintSemicolon
\KwInput{RDF Triples, no. of fragments}
\KwOutput{list of frequent subjects}
\SetKwProg{Fn}{Function}{:}{}
  \Fn{\FEXT{RDF Triples,k fragments}}{
  hashmap h with key = subject and value = frequency\\
       \For{each triple in triples}{h(subject)++}
       sort the hashmap by value\\
    \textbf{return} top k subjects
  }
\label{alg:a1}
 \caption{Extraction of popular Nodes}
\end{algorithm}

In algorithm 2 after getting the most important subjects, we allocate the triples corresponding to that subject to a cluster. We then obtain k fragments. To allocate the remaining triples to these fragments, we need to find out the degree of closeness of each triple with the existing fragments. Given a triple, t not yet assigned to any fragment, we find out which fragment has the most number triples which contain the object equivalent to the subject of triple, t. The triple t and all other triples which share the same subject which we call the secondary master node are then added to that cluster. This method is continued for the rest of the triples.  Here each of the triple not part of initial partitioning is compared with every existing triple of the partitions. We do the preprocessing of partitions and use HashMap to record the occurrences of objects in a partition. Since every subject of the remaining triple is checked in the HashMap, the worst time complexity of the algorithm becomes O(n*k) where n is the number of triples and k is the number of fragments or partitions. 

\begin{algorithm}[H]
\SetKwFunction{FSAP}{SAPartition}
\DontPrintSemicolon
\KwInput{RDF Triples, frequent subjects,k fragments}
\KwOutput{k partitions}
\SetKwProg{Fn}{Function}{:}{}
  \Fn{\FSAP{RDF Triples, frequent subjects,k fragments}}{
       \For{each fragment f=1 to k}{fragment fi = matches of triples with subject si}
       make hashmap h for each fragment i where key = object and value = frequency
       \For{rest of the triples}{put triple in fragment with maxhi(object)}
       \textbf{return} k fragments
  }
\label{alg:algo2}
 \caption{Semantic Aware Partitioning}
\end{algorithm}

\subsection{Distribution using Partial Replication}

When the user submits a query to the coordinator node, it will be answered using graph traversal from all the available nodes in the distributed environment in LBSD. This section includes details of the replication and distribution strategy.

After the fragmentation of the dataset, it is not necessary that we get fragments that are almost equal in size primarily because the frequency of outgoing edges is not uniformly distributed in the triples. While some nodes might have a high number of outgoing edges, others might barely have that many outgoing edges. This might lead to skewed distribution, which will result in unequal load distribution and delayed query execution time. To mitigate this problem, we calculate the sizes of the fragments and allocate them to different sites in such a way that there is an approximately equal load at each of the sites. So, a fragment of bigger size should be placed with a fragment of smaller size.

\begin{algorithm}[H]
\SetKwFunction{ALL}{Allocation}
\DontPrintSemicolon
\KwInput{fragments}
\KwOutput{clusters}
\SetKwProg{Fn}{Function}{:}{}
  \Fn{\ALL{fragments}}{
  Array A of {fragment, size of fragment}
       \For{every fragment f do}{A.insert(F,sizeof(F)}
       sort A in descending order of size
       \For{every index i in A do}{Add A[i].fragment to cluster with lowest size}
       
  }
\label{alg:algo3}
 \caption{Distribution of Clusters}
\end{algorithm}

 For Replication, LBSD uses centrality measurement. Degree centrality measures the number of incoming and outgoing relationships from a node. The Degree Centrality algorithm can help us find popular patterns(subject-object) in a graph. This is built-in feature of Neo4j\cite{neo4j}. For each predicate, we can measure the centrality that how much they are connected to subjects. Having the same centrality predicates should lie on the same host in a distributed environment. Here for our experiment, centrality lies between interval [0.14,1]. The highest centrality is 1, when the count of a number of subjects and number of labeled edges become same. Centrality helps to replicate and distribute data among available nodes.  

\begin{algorithm}[H]
\DontPrintSemicolon
\KwInput{Triples}
\KwOutput{Replicated Data}
 
 \For{each cluster c=1 to n}    
        { 
        \For{each Predicate p}{\If{$ cen(p) > Max[Ap]$ \tcp*{MAx[Ap]= centrality of max.occurring predicate}}
    {
        \For{i=1 to n}{Pi =p \\Data will be replicated on node for cluster c[i=1 to n]}   
       
    }}
        	
        }
\label{alg:algor3}
 \caption{Replication}
\end{algorithm}

Replication replicates the data to the available nodes in the distributed system. Partial replication only replicates a few amounts of data that satisfy the given threshold value or cut-off. Here we have frequent patterns and its centrality.  According to top k subjects analysed from algorithm 1, will have top k patterns. That means properties associated with those subjects. These top k patterns help to decide the replication level. So, the centrality of the top pattern becomes the threshold value for partial replication which is known as Max. [Ap]. For example, some subject k1 is there in list of k subjects having centrality 0.58, then patterns of centrality between 0.58 to 1 will be replicated . Here in LBSD it is 0.65 i.e. patterns' centrality between 0.65 to 1 were replicated counted as most frequent one.

\section{Experimental Details}
The hardware setup consists of Intel® Core (TM) i3-2100 CPU@ 3.10GHz 3.10 GHz 8GB. The software setup consists of Neo4j Desktop 1.1.10 \cite{neo4j} and for visualization neo4j browser version 3.2.19 is used. We have used NeoSemantics\cite{neosemantics} to upload rdf supported data files. As a distributed database we have used DGraph v1.0.13 \cite{dgraph}. 
\subsection{Benchmark Dataset and Queryset}
Linked Observation Data \cite{lod} (LOD) benchmark dataset is used for the experiment. LOD has near about 3000 categories. From that, we have used Linked Sensor Data(LSD) comprises of results of sensor observation results of Hurricane in the US. The dataset includes different observations of Wind Direction, Wind Gust, Air temperature, Humidity, Precipitation, etc. 

The benchmark LSD query set is used for the experiments. It consists of 12 queries which are classified into four types. Type 1 and Type 2 queries are linear and star queries respectively. Type 3 and Type 4 queries are Administrative or Range queries and Snowflake queries respectively.There are 3, 4,3, and 2 queries of Type 1 ,2, 3,and 4 respectively. Linear queries select some predicates from data and Star queries select specific subject/objects relevant to the given predicate. Administrative or Range queries are used to retrieve data using aggregation function or range function and Snowflake queries are a combination of both Type 1 and Type 2. 

\subsection{Evaluation Parameters}
Performance of LBSD will be evaluated using the following quantitative and qualitative evaluation parameters:
\subsubsection{Quantitative parameters} 
This section discusses quantitative parameters that measure the performance of LBSD in terms of some percentage or value.\\
\textit{Algorithm Execution time (AET)}is the time taken by the execution of all three algorithms of LBSD.\\
\textit{Inter-Node Communication (INC)} is measured in terms of how much communication cost is there to answer a query using different nodes.\\
\textit{Query Execution Time (QET)} is the time taken by a query to complete execution.\\
\textit{Query Join (QJ)} measures the number of join operations to execute a query.

\subsubsection{Qualitative parameters} 
This section discusses qualitative parameters which compare the LBSD in terms of quality measures.\\
\textit{Partitioning technique} defines the technique used for the partitioning of data.\\
\textit{Distribution technique} defines the technique used for the distribution of the RDF graph.\\
\textit{Workload information} informs that is there any query workload information required for the execution.\\
\textit{Replication strategy} defines the technique to replicate partitioned data.\\
\textit{Scalability} defines how the system reacts when the data size increases.
\textit{Storage Requirement} gives an idea about amount of storage space used by system.

\section{Results and Discussion}

LBSD is demonstrated using LSD benchmark data and query set. This section presents results for basic and scaled query execution time, Algorithm execution time. It also contains discussions about the choice of replication level. The results for other quantitative parameters like query joins and INC are also included here. 

\subsection{Basic Query Execution Time (QET)}
QET analysis for LBSD has been done for LSD. There are four types of queries and results are taken by analyzing performance for each of them. QETs are averaged over three consecutive executions to reduce fluctuations for each query. Further all the QETs are averaged over all the queries of that type.
\begin{figure}[h!]
    \centering
    \includegraphics[width=0.6\columnwidth]{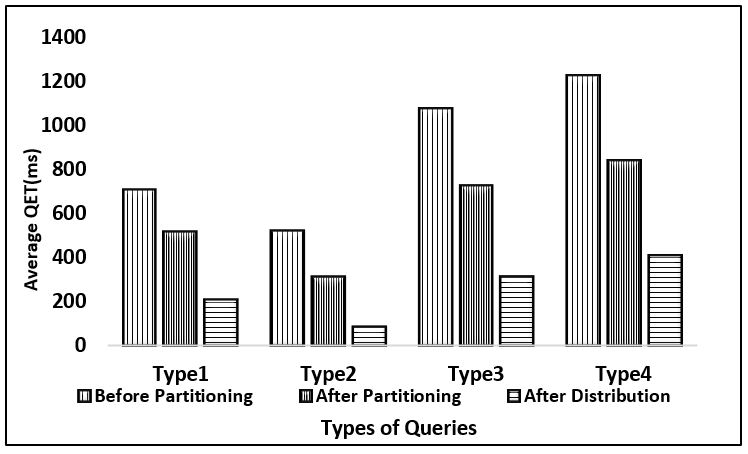}
    \caption{Average QET for all types}
    \label{fig:qet1}
\end{figure}

Figure \ref{fig:qet1} shows that Type 2 queries are taking less amount of time because it is just fetching the values whereas Type 4 queries are taking a larger amount of time compared to all types of queries.

\subsubsection{Data Scaling for QET}

\begin{figure}[h!]
    \centering
    \includegraphics[width=0.6\columnwidth]{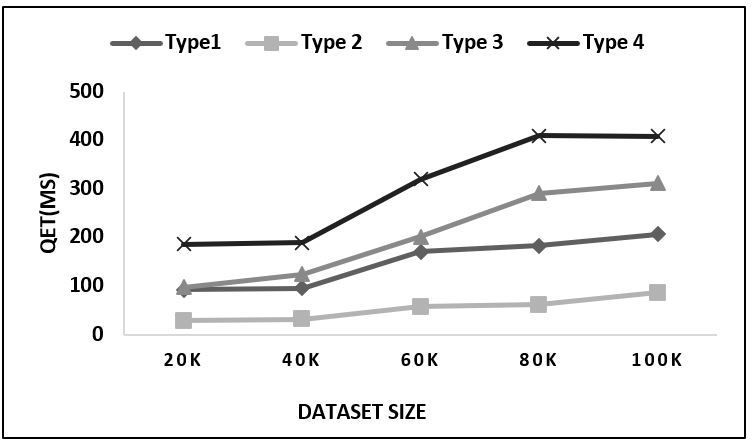}
    \caption{QET Before Partitioning}
    \label{fig:qet2}
\end{figure}

Data scaling experiment done for the size 20k to 100k. Figure \ref{fig:qet2} shows that QET increases with increase in the datasize from 20k till 100k for all the query types. This increase is more pronounced for Type 2 and Type 3 queries. 

\begin{figure}[h!]
    \centering
    \includegraphics[width=0.6\columnwidth]{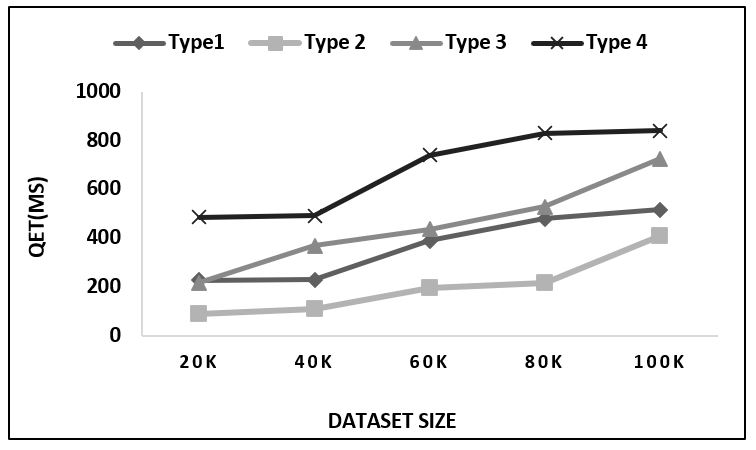}
    \caption{QET After Partitioning}
    \label{fig:qet3}
\end{figure}

Type 2 queries taking a large amount of time when data size increases 40k to 60k as the value required to fetch is distributed over nodes. For all types of queries as data size increases QET is getting reduced after distribution as shown in figure \ref{fig:qet3}. There is on an average 71\% of QET gain for all types of queries. 

\subsection{Algorithm Execution Time (AET)}   
\begin{figure}[h!]
    \centering
    \includegraphics[width=0.6\columnwidth]{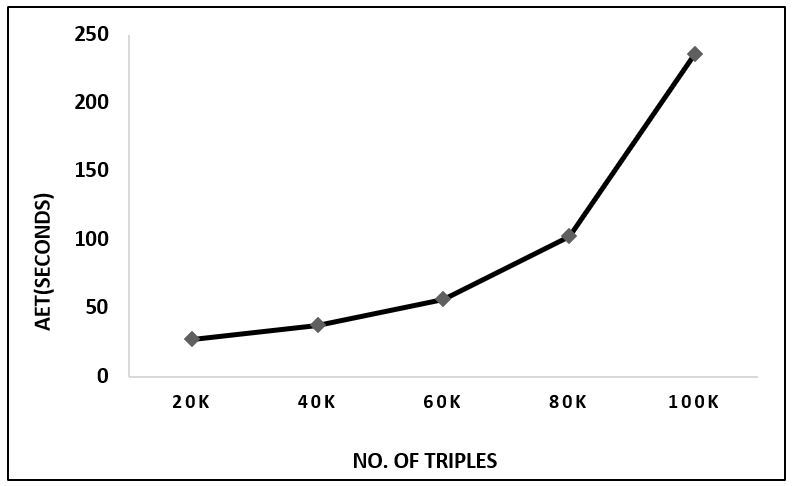}
    \caption{Algorithm Execution Time}
    \label{fig:aet}
\end{figure}

There are three algorithms used by the LBSD system. Algorithm 1 and Algorithm 2 are used in first phase and second phase of LBSD uses Algorithm 3 and Algorithm 4. The total execution time taken by the system to execute all four algorithms for different data sizes is shown in figure \ref{fig:aet}. We can see that as data size increases AET increases. There is a ramp shown in the graph when data size increases from 80k to 100k.

\subsection{Replication level}              
\begin{figure}[h!]
    \centering
    \includegraphics[width=0.6\columnwidth]{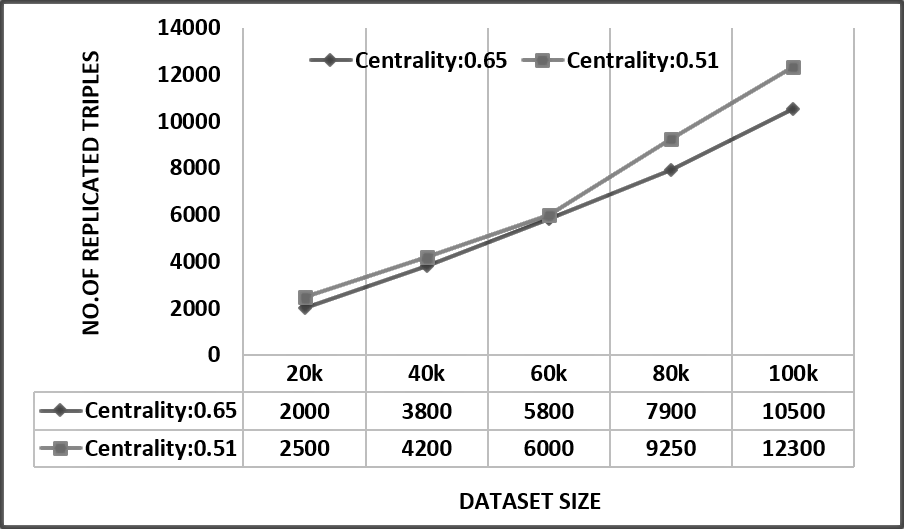}
    \caption{Replication level for centrality 0.65 and 0.51}
    \label{fig:replicationlevel}
\end{figure}

For partial replication, to decide replication level first we have kept threshold at centrality 0.65.As shown in figure \ref{fig:replicationlevel} there is a linear increment in no. of triples to be replicated with increasing data size. On average 12\% of data were replicated. When we have changed the threshold value to 0.51, no. of triples increased with an average of 14\% data were replicated. But for this experiment, It has been found that centrality 0.65 is optimal.  

\subsection{Query Joins}
If we compare LBSD to DWAHP\cite{dwahp} or to any such relational system, it works better in terms of Query Joins (QJ). In the graph database, we can access the whole database by traversing an edge, which reflects the absence of QJ. This is an advantage of LBSD that it eliminates QJ for accelerating queries over distributed data. 

\subsection{Inter Node Communication}
Inter Node Communication (INC) means the amount of communication requires between available nodes in a distributed environment. 

\begin{figure}[h!]
    \centering
    \includegraphics[width=0.6\columnwidth]{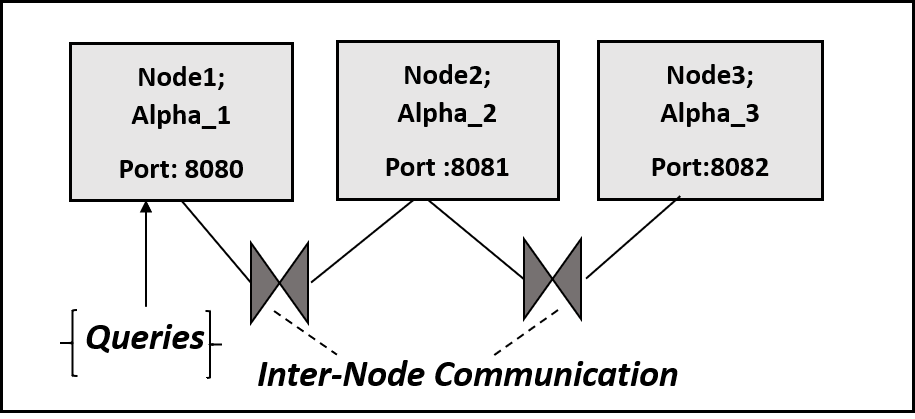}
    \caption{Inter Node Communication}
    \label{fig:inc}
\end{figure}
As shown in figure \ref{fig:inc}, INC is defined by communication cost that query uses to answer using three available nodes, named as; Alpha\_1, Alpha\_2, Alpha\_3. These three nodes are working on a different port of DGraph Ratel. While accelerating queries over distributed data, out of 12 queries on an average of 7 queries are answered by the local node. So approximately 58\% of queries are answered without INC.

\subsection{Comparison of LBSD with the state of the  art}

LBSD is compared with different techniques including APR\cite{lenawiesecnf} which is graph based state of the  art work and DWAHP\cite{dwahp} , similar relational state of the  art work. This section describes the detailed comparison of qualitative and quantitative parameters listed in Section 4.2. 

 The table \ref{quality} shows the qualitative comparison of LBSD with APR and DWAHP.LBSD is a semantic aware partitioning technique whereas APR uses METIS \cite{metis}, a simple weighted partitioning technique. Semantic aware partitioning technique keeps semantic relation between nodes alive even after partitioning. Due to semantic relation, similar nodes will be on the same host which will accelerate linear and star queries. DWAHP is a relational approach which uses hybrid partitioning technique using a combination of property and binary tables. For the distribution, LBSD uses Centrality replication threshold whereas APR uses Global Query graph approach to identify border nodes communication cost. It helps to reduce communication cost between nodes on different hosts through replication. LBSD is a static partitioning approach and APR is an adaptive approach for distribution and partitioning. LBSD does not require workload information of the implementation whereas APR and DWAHP both requires workload information. For adhoc queries, workload aware system needs to re-run the algorithm for updated workload information. APR is implemented in such a way that it also takes space in consideration while distributing data. APR works well with storage adaption at three levels. It also uses compressed replication technique in which it compresses long URI to numerical value. DWAHP and LBSD use Partial replication technique. APR and DWAHP exhibit better scalability compared to LBSD.

\begin{table*}[h!]
\centering
\caption{Qualitative Comparison of LBSD with state of the art}
\begin{tabular}{||m{4cm}|m{3.5cm}|m{3.5cm}|m{3.5cm}||} \hline \hline
Parameters& LBSD & DWAHP \cite{dwahp} & APR \cite{lenawiesecnf}   \\ \hline \hline
Partitioning technique & Semantic Aware & Hybrid & Simple \\ \hline 
Distribution technique &  load balanced, Using replication Threshold   &  Not load balanced,  Using n-hop reachability matrix  &   load balanced, Using Global Query Graph \\ \hline
Workload information &	Not required &	Required &	Required \\ \hline
Replication technique &	Partial replication & Partial Replication &	Compressed replication \\ \hline
Scalability & Poor beyond 4x &	Better & Better \\ \hline 
Storage Requirement & Basic data + 12\% replicated data & Basic Data + around 20\%  replicated data&  Basic Data +30\% replicated data RDF 3X indexing \\ \hline \hline
\end{tabular}
\label{quality}
\end{table*}

The quantitative results for LBSD are shown in table \ref{quantity}. The average QET gain for all types of Query reported 71\%. The AET is averaged over all four algorithms and its total time complexity is O(n). The INC is  58\%, i.e. 58\% queries are answered without INC. LBSD eliminated complex query join operations.

\begin{table*}[h!]
\centering
\caption{LBSD Results}
\begin{tabular}{||c|c||} \hline \hline
Parameters& LBSD \\ \hline \hline
QET & 71\%  \\
AET & O(n)  \\
INC & 58\%  \\
QJ & eliminated \\ \hline \hline
\end{tabular}
\label{quantity}
\end{table*}

QET for DWAHP and LBSD is almost the same. LBSD is reporting faster AET compared to DWAHP but it shows poor scalability beyond 4x. LBSD and APR being graph-based techniques, are able to eliminate Query joins. INC is approximately the same for both LBSD and DWAHP. APR generates small number of big clusters as compared to LBSD which reduces INC for APR.As a result of it, LBSD and DWAHP queries need to scan lesser data.APR replicates almost 30\% of data whereas APR needs to replicate only 12\% of data. As APR needs to rerun the algorithm for updated workload , the AET of LBSD reports faster than APR. There is an indexing overhead, as APR uses RDF-3X engine which requires indexing over all the three columns Subject, Object, and Predicate.

\section{Conclusion}
This method implemented to manage the increasing size of RDF data management by semantic aware partitioning and distribution of data using graph approach. Based on in-degree and out-degree of vertices LBSD partitions the data. For distribution purposes, we have distributed data on available three virtual nodes. LBSD compared in terms of two types of parameters: Qualitative and Quantitative. To analyze performance in terms of QET, the system uses 4 types of queries. It shows an average 71\% gain for all types of queries after distribution. QET gain for type 2 queries in scalability experiments increases linearly with an average gain of 72\% as it has lower INC whereas type 4 has an average gain of 55\% as data size increases from 20k to 100k. The system also shows better performance in terms of inter-node communication as it answers 58\% of the query by the local node. The scalability results show that AET increases rapidly when data size increases from 80k to 100k. We can make this system adaptive to deal with dynamic data in the future.

\section{Acknowledgments}

We would like to thank Dr.Trupti Padiya, Postdoctoral Researcher, Friedrich Schiller University Jena for helping us to resolve technical issue during this research.

\bibliographystyle{abbrv}
\bibliography{Ideas}  
%

\end{document}